# Automatic elimination of the pectoral muscle in mammograms based on anatomical features

Jairo A. Ayala-Godoy[1], Rosa E. Lillo[2] and Juan Romo[3]

[1] Department of Basic Sciences and Engineering, Universidad del Caribe,
Cancun, Quintana Roo, Mexico

[2] Department of Statistics and UC3M-Santander Big Data Institute, Universidad Carlos III de Madrid,
Madrid, Spain.

[3] Department of Statistics and UC3M-Santander Big Data Institute, Universidad Carlos III de Madrid,
Madrid, Spain.

**Abstract**

Digital mammogram inspection is the most popular technique for early detection of abnormalities in human breast tissue. When mammograms are analyzed through a computational method, the presence of the pectoral muscle might affect the results of breast lesions detection. This problem is particularly evident in the mediolateral oblique view (MLO), where pectoral muscle occupies a large part of the mammography. Therefore, identifying and eliminating the pectoral muscle are essential steps for improving the automatic discrimination of breast tissue. In this paper, we propose an approach based on anatomical features to tackle this problem. Our method consists of two steps: (1) a process to remove the noisy elements such as labels, markers, scratches and wedges, and (2) application of an intensity transformation based on the Beta distribution. The novel methodology is tested with 322 digital mammograms from the Mammographic Image Analysis Society (mini-MIAS) database and with a set of 84 mammograms for which the area normalized error was previously calculated. The results show a very good performance of the method.

## 1. Introduction

According to the World Health Organization (WHO) [1], Breast cancer is the most frequent cancer among women, impacting 2.1 million women each year, and being the leading cause of cancer death among women. WHO also notes that in 2018, it was estimated that 627,000 women died from breast cancer. At present, there are no effective mechanisms for prevent it because its cause is unknown. However, it is well established that a diagnosis in its initial stage offers very good prospects to be treated successfully [2].

Mammography is a flat image of the breast obtained through X-rays, which plays a very important role since it is currently the most widely used method for the analysis of breast cancer [3]. However, an accurate diagnosis is still a challenging clinical task for radiologists ( [4]) due to diverse factors, such as: high heterogeneity (size, shape, texture, or color), variability in the appearance of the abnormalities, characteristics of the breast tissue and the quality of the images. For this reason, some automatic methods have been proposed in the literature with the objective of reducing clinical analysis errors and radiologist discrepancies, although all these methods are still quite improvable with respect to the final output.

Kowsalya *et al.* [5] commented on the difficulty of interpreting digital mammograms without a preprocessing phase, so they recommend performing a processing stage before applying any analysis to the mammogram, since it is essential to find the edges of the region of interest without deviations from background. This problem is particularly evident in the mediolateral oblique view (MLO) where labels, wedges and pectoral muscle could disrupt focus on the breast tissue. Failures in preprocessing may lead to important bias in the analysis [6].

Several approaches to automatic elimination of the pectoral muscle have been proposed. The mini MIAS [7] database is the most used dataset, and only a few approaches evaluate all the mammograms in this database [8]. Ferrari *et al.* [9] used the Hough transform and Gabor wavelets. Kwok *et al.* [10] proposed a method based on the upon cliff detection. Ma *et al.* [11] identified the pectoral muscle through adaptive pyramids and minimum spanning trees. Kinoshita *et al.* [12] proposed a straight line method based on radon domain information. Yuan *et al.* [13] presented a method based on Markov chain and active contour model. *Zhou et al.* [14] developed a texture-field orientation method that combines a priori knowledge and local and global information. Camilus *et al.* [15] used a graph cut-based image segmentation technique. Chakraborty *et al.* [16] presented a method based on average gradient and shape features. Nagi *et al.* [17] used a detection method with morphological pre-processing. Chen *et al.* [18] presented a region growing method with the seed point located close to the border of the pectoral muscle and the breast tissue. Maitra *et al.* [6] also presented a region growing method with three steps: contrast enhancement, defining the rectangle to isolate the pectoral muscle region. Czaplicka *et al.* [19] used multilevel Otsu threshold and refinement of initial segmentation by linear regression. Liu *et al.* [20] presented an algorithm using Otsu thresholds and multiple regression analysis. Li *et al.* [21] segmented the pectoral edge combining characteristics of homogeneous texture and intensity deviation, refining the procedure using the Kalman filter. Oliver *et al.* [22] used intensity and texture information in the probabilistic model to segment a mammogram. Liu *et al.* [23] proposed a method which utilizes statistical features of pixel responses. Chen *et al.* [24] proposed a shape-based detection method for extracting the boundary of the pectoral muscle in mammograms. Sreedevi *et al.* [25] proposed global thresholding in combination with edge detection and connected labeling technique. Vikhe *et al.* [26] used a thresholding method based on intensity for pectoral muscle boundary detection. Yoon *et al.* [27] presented a thresholding method with morphological operations and random sample consensus algorithm. Yuksel *et al.* [28] presented novel multistage scheme for pectoral muscle removal from mammography images.

In this paper, a method based on two important anatomical features for the automatic elimination of the pectoral muscle is proposed. These important anatomical features are: the pectoral muscle has almost homogeneous gray level values ( [9]) and there is a significant difference in the intensity variation between the pectoral muscle and the breast tissue [13].

We will refer to this method from now on as Automatic elimination of the pectoral muscle in mammograms based on anatomical features (AEPm). It is important to highlight that our method takes into account all the different characteristics that can have each mammogram (size, curvature, texture, gray levels, among others). Since the precise elimination depends largely on finding the edge of the muscle, our proposal is based on the use of intensity transformations, which is a useful tool to improve the detection of the true edge.

The method AEPm has two steps: (1) a procedure to remove the noise of the image, deleting elements such as labels, markers, scratches, wedges, etc. This step is carried out through a histogram thresholding automatic technique, which is an excellent way to extract useful information encoded in pixels, minimizing background noise [29]. In this step an automatic threshold is chosen in each mammogram, specifically where the first local minimum of the gray level empirical density appears. The idea is to separate the image into two parts; the background and region of interest. (2) An intensity transformation based on the Beta distribution function is applied to the processed image in step 1. This transformation is the novel methodology proposed in this article, which takes into account the knowledge about the two anatomical features of the breast mentioned above: increasing the intensity of the pectoral muscle and decreasing the intensity of the regions near the muscle with low intensities.

This paper is organized as follows. Section 2 describes the two steps used to eliminate the pectoral muscle: the histogram thresholding technique for removing the background of the image and the intensity transformations based on the Beta distribution function to identify the pectoral muscle. In Section 3 the results obtained using the AEPm method introduced in this paper and others used in the literature are discussed after to be applied to two real data sets. Finally, the main conclusions are summarized in Section 4.

## 2. Methodology

There are two important regions in a mammography: the pectoral muscle and the breast region. In the MLO view, the pectoral muscle is located in the top, contained in a right triangle, (because the imaging is performed with the patient in a standing position, forming a right angle) and the inferior region of the breast, where the breast tissue can be approximated as an oval, (see Figure 1).

Therefore, each input image has the pectoral muscle at the top left; that is, all the images where the breast appears on the right side are reflected vertically. The origin of the coordinate system is at the top left corner of the image, where $x$ is defined as the horizontal axis and $y$ as the vertical axis, $x = 1, \dots, n_x = 1024$ and $y = 1, \dots, n_y = 1024$. The intensity level of each pixel is denoted by $I(x, y)$, where $0 \leq I(x, y) \leq 1$. To eliminate the pectoral muscle, the AEPm method consists of two main steps: (1) remove the image noise and (2) identify of the muscle edge.

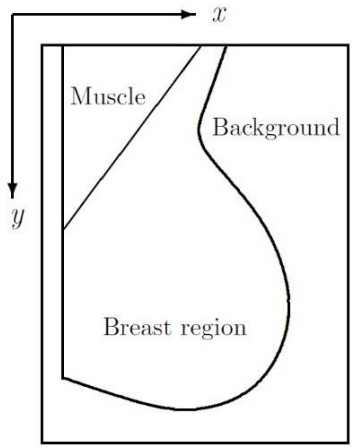

Figure 1: Representation of a mammography.

## 2.1 Removing the image noise

Removing the image noise plays a crucial role in the field of medical imaging in terms of improving image quality, where each image may contain many objects, such as labels, markers, scratches, adhesive tape, that need to be removed [30]. Figure 2 shows an example of an image that contains some background objects. To remove these objects we used a threshold technique based on empirical density of the gray level of the image [31]. This method is used to separate specific regions from the rest of the image.

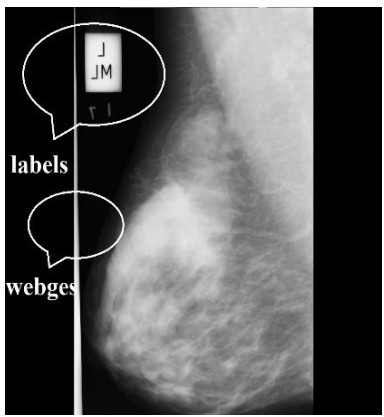

Figure 2: Example of a mammogram which contains some background labels and wedges, image mdb115 in mini-MIAS.

The idea is from a gray-scale image, create a binary image that help us to characterize all the absurd objects in the image and, thus, be able to eliminate the noise elements from the background of the image [32]. The key parameter in the process is the choice of the threshold value in the gray scale [33]. In the AEPm, this parameter is chosen independently for each mammography automatically detecting when a significant change of the gray levels between the background and the breast region occurs. The first step of the AEPm method can be summarized in the following four phases:

1. Plotting the empirical density of the gray level of the image, the threshold (denoted by $c$) is found as the first local minimum of the empirical density. An illustration is shown in Figure 3. The value $c$ is computed automatically for each image.

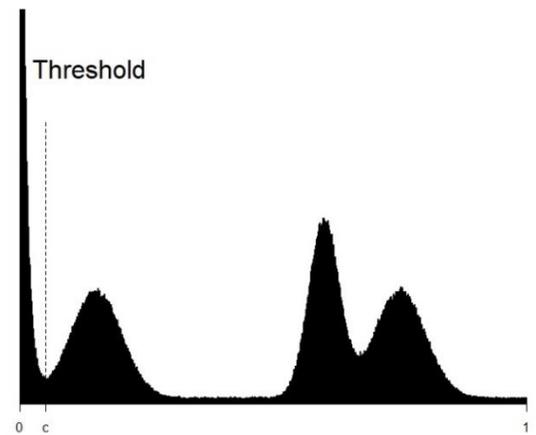

Figure 3: Obtaining $c$

2. Applying the function,

$$D(x,y) = \begin{cases} 1 & \text{if } I(x,y) > c \\ 0 & \text{if } I(x,y) \leq c, \end{cases} \quad (1)$$

the initial image is transformed in a binary image, (see Figure 4(b)).

3. The matrix obtained from the binary image is used as an input element in the function "bwlabel" of the EBImage package in **R** [34]. This function extracts every connected sets of non-zero pixels from an image and relabel these sets with an increasing integer, where the pixels labeled 0 are the background, the pixels labeled 1 make up one object, the pixels labeled 2 make up a second object, and so on. Hence, to obtain the region of interest, where it is found breast region and pectoral muscle, we take the larger object and eliminate the other objects, (see Figure 4(c)).

4. Finally, a clean image is obtained by replacing all the pixels with value 1 from the binary image by their initial value, (see Figure 4(d)).}

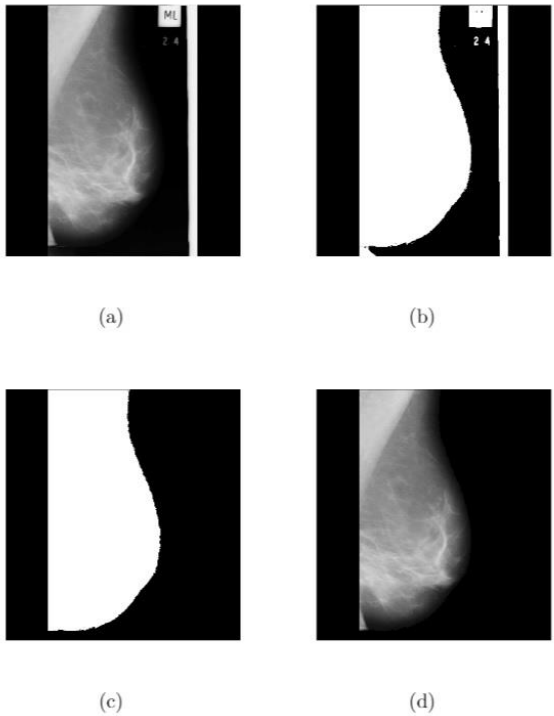

Figure 4: Example of how to remove the image background, image mdb051 in mini-MIAS. (a) Mammogram oriented such that the pectoral muscle is located at top left side. (b) The mammogram converted to a binary image. (c) Small objects removed from the image and considering only the large object. (d) Original pixel values of the mammograms are returned without noise in the background.

With these four phases, the first step of the AEPm method is completed.

## 2.2 Identification of the pectoral muscle

The pectoral muscle represents a region of predominant density in the MLO view of mammograms. Its inclusion affects the results of intensity-based imaging methods or bias procedures in breast cancer detection. Therefore, removing the pectoral muscle accurately from a mammogram is an essential component [35]. To identify the pectoral region we used an intensity transformation for image enhancement [36]. This technique is based on spatial operations, which are performed directly on the pixels of a given image. The transformation function $T$ is of the form:

$$g(x,y) = T[I(x,y)], \quad (2)$$

where $I(x,y)$ the gray level intensity of the input image is, $g(x,y)$ is the level gray of the output (processed) image, and $T$ is an operator on $I$ defined over the coordinates $(x,y)$.

In the AEPm method, the *Beta distribution function* is considered in equation (2). After performing several empirical tests with the classic intensity transformations ( [37]) and taking into account the high heterogeneity of mammograms (size, shape, present anomalies, contrast, intensity, among others), We decided to use the Beta distribution function as an intensity transformation which represents a novel methodology, because thanks to its two parameters we can obtained a great variety of ways to transform each image. With the aim of the beta distribution function as intensity transformation, we take advantage of the two anatomical features mentioned in Section 1 and thus two objective are achieved: highlight the entire pectoral muscle (the pectoral muscle has almost homogeneous gray values [9]) and separate the pectoral muscle from the breast tissue (there is a significant difference in the intensity variation between the pectoral muscle and the breast tissue [13]). At this point, our work focused on automatically tuning the parameters of the Beta distribution function to find good results, depending on the intensity characteristics of each mammogram,

$$g(x,y) = BDF[I(x,y); \alpha, \beta]$$
$$= \frac{1}{B(\alpha,\beta)} \int_0^{I(x,y)} t^{\alpha-1}(1-t)^{\beta-1} \, dt \quad (3)$$

As it is known, the Beta distribution function is very versatile and can adopt different shapes depending on the values of its parameters. Then, the parameters can be adjusted to highlight the pixels where the pectoral muscle is located depending on the gray scale of each mammogram. Both parameters α and β in (equation (3)) are stretching parameters that help us to find the adequate transformation on each mammogram. This transform is used to compress or expand the intensity level for each pixel.

Therefore, based mainly on the shape of the function $g$ and after performing an exhaustive exploratory analysis, we have decided to fix the parameter $\alpha = 5$ and tuning the parameter β in the interval [2,6]. In Figure 5(a) we can see how the shape of the Beta distribution function varies with these parameters. Also, we can see an example in the Figure 5(b), 5(c), 5(d), where we transformed a mammogram with three different values, $\beta = 2,4,6$ respectively. The second step of the AEPm method consists of the following seven phases:

1. The final image of step 1 is the starting point of step 2.
2. An intensity transformation with the Beta distribution function is applied to each pixel, that is, $T[I(x,y)] = BDF[I(x,y); 5, \beta]$, where the tuning parameter β takes values $\wp = \{2 + 0.1i, i = 0, ..., 40\}$. For each mammogram we automatically choose the value of β with better results in phase 6.

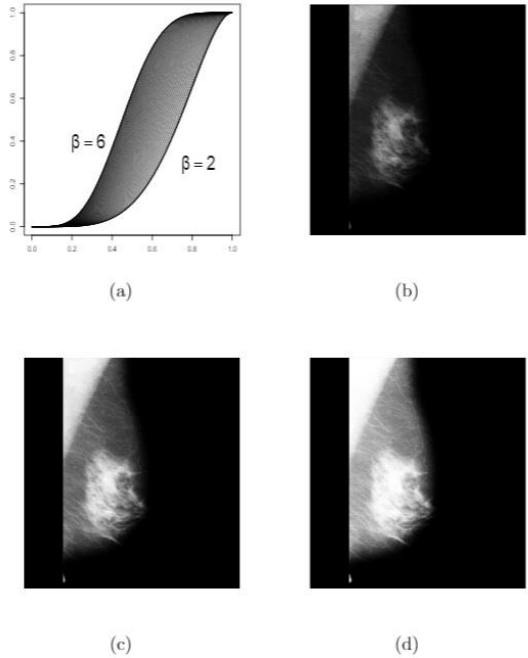

Figure 5: (a) Possible shapes of the $BDF[I(x,y); 5, \beta]$ transformation function. In (b), (c) and (d) we applied the transformations $BDF[I(x,y); 5, \beta]$, $\beta = 2,4,6$ respectively, image mdb042 in mini-MIAS.

3. For each image $k$, the following index is calculated

$$\mu_k = \frac{1}{m} \sum_{x=1}^{n_x} \sum_{y=1}^{n_y} I(x,y), \quad (4)$$

where $m$ is the number of pixels with non-zero intensity of image $k$.

4. For each image $k$ and each β, we selected in each row from left to right the first coordinate non-zero $(x, y)$ that is less than $\mu_k$. The selected points were connected and form the rough initial edge for each value of β. In the Figure 7(b) we can see an example of a rough edge with the adjusted parameter β.

5. Now, it is necessary to refine these rough edges in order to better approximate the real edge. Therefore, we smoothed the edges of the previous phase by means of a cubic B-spline, (see Figure 7(c)). The coordinates of the edges for each β value will be denoted as:

$$(x_{\beta,1}^k, 1), (x_{\beta,2}^k, 2), \ldots (x_{\beta,n_{\beta,k}}^k, n_{\beta,k})$$

where $n_{\beta,k}$ will be the size of each edge.

6. Taking into account the two anatomical features mentioned before, that are, the pectoral muscle has almost homogeneous gray level values ( [9]) and there is a significant difference in the intensity variation between the pectoral muscle and the breast tissue [13], we designed the following criterion to choose automatically the value of β for each image:

$$\widehat{\beta_k} = \max_{\beta \in \wp} |g(x_{\beta,y}^k - 2, y) - g(x_{\beta,y}^k + 2, y)| \quad (5)$$

In Figure 6 we show an example of the different edges obtained in the process to find the optimal beta value of the equation (5) where the red edge is the edge provided by the optimal β.

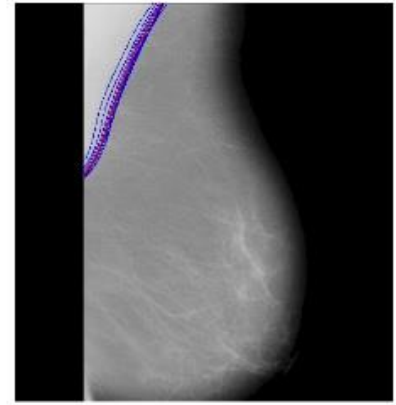

Figure 6: Different edges obtained in the process to find the optimum beta value of the (equation 5), image mdb006 in mini-MIAS.

7. Finally, after estimating the muscle edge, the region located to the left of the edge is eliminated, which should be the pectoral muscle region, (see Figure 7(d)).

The AEPm method was implemented in the programming language **R** version 3.6.1. The algorithm uses simple functions, which in the worst case must pass through the complete matrix. Therefore, the complexity of the algorithm is the order $O(n^2)$, where $n$ is the index of the image dimension $n \times n$ pixels. To analyze the execution time of the algorithm, we take a mammogram and run the algorithm with different sizes. The sizes run from $100 \times 100$ to $2000 \times 2000$ pixels. We calculate the execution times of 100 replicas of each size and the results can be seen in Figure 8.

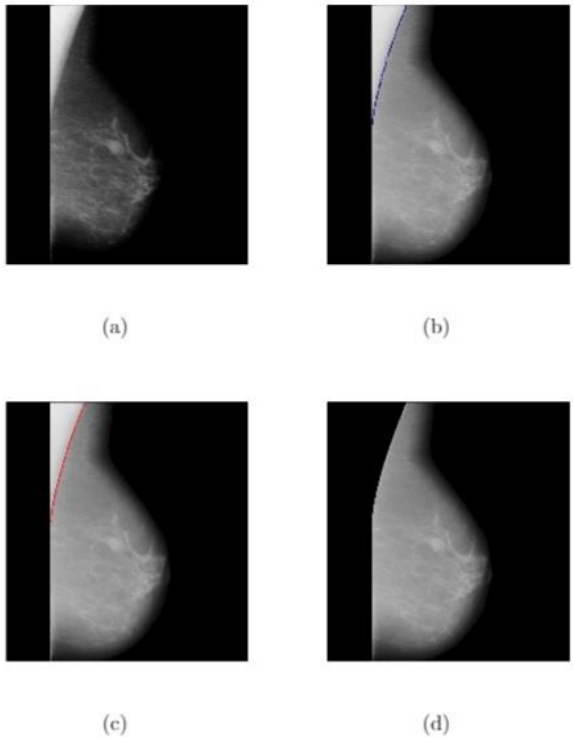

Figure 7: Example of how to carry out the second step of the AEPm method, image mdb012 in mini-MIAS. (a) Beta transformation is applied. (b) Rough edge with the adjusted parameter $\beta$ in blue. (c) Smoothed edge by means of a cubic B-Spline in red. (d) The eliminated pectoral muscle.

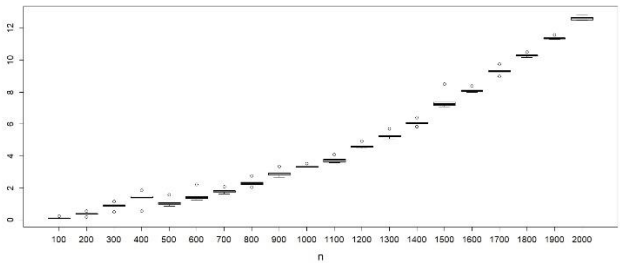

Figure 8: Executions times of the AEPm method

Once the AEPm has been explained, in the next section its performance will be shown in comparison with other methods already used in the literature.

## 3. Results and discussion

Beta distribution was used as an intensity transformation which provides an effective method to identify the pectoral muscle edge. The parameter β plays an important role, since it is adjusted in each image to optimize the detection of the real muscle edge. Several studies to analyze the behavior of β were carried out and we consider that the way we adjusted it was quite adequate, since this parameter captures the existing heterogeneity in each mammography. We could observe, as an empirical result that when the pectoral muscle has high levels of intensity, the value of β is small and vice versa. Moreover, we found that when the breast tissue is very close or superimposed over the edge of the muscle, the adjustment of the parameter is quite sensitive.

The AEPm method was tested with the mini-MIAS database. This is a set of mammograms gathered by Mammographic Image Analysis Society (MIAS), organization from UK [7]. It contains 322 digitized mammograms of 200 $\mu$m/pixel, 8 bits/pixel and 1024 × 1024 pixel size. The database includes MLO views of both left and right breasts from the same patient, the classification of the type of tissue, including presence and location of abnormalities and the classification of the severity type. These annotations were developed by experts on the subject through clinical studies.

We use two well known methods to evaluate the performance of our proposal: *acceptable rate* and the *area normalized error*, which are explained below. We compare the results obtained with previous works that used the same evaluation mechanics and the same database.

### 3.1 Acceptable rate

Each identified edge is classified in three categories (of the five categories shown in [10]):

**Exact**: The identified edge fits the pectoral margin exactly. Any deviation from the visually perceived margin are imperceptible or insignificant.
**Adequate**: The line found identify the pectoral muscle margin inexactly, but with sufficient accuracy to localize the pectoral margin.
**Inadequate**: The identified edge was not found or was inaccurate in localizing the pectoral muscle edge.

The *acceptable rate* is conformed by the categories exact and adequate. For this method, we evaluated the 322 mammograms from the mini-MIAS database. In Table 1, the acceptable rates is calculated, obtaining a value of 95.34% for the AEPm method.

In Table 2, the comparison between the AEPm method and the previous studies are displayed. A better performance of the AEPm method can be appreciated.

The AEPm method shows versatility to adapt to the different anatomical characteristics of the mammograms sizes, curvatures, textures, gray levels, etc.). For instance, the AEPm method correctly recognized when the pectoral muscle was small (Figure 9(a)) or large (Figure 9(b)), even when the difference of gray levels between the pectoral muscle and its surroundings was low, as is shown in Figure 9(c). In addition, in some cases, the compression of the breast when the mammography is taken can introduce noises that generate falsely recognized pectoral muscle edges. However, in these cases, the method correctly recognized the edges as is shown in Figure 9(d).

Table 1: Acceptable rate for the AEPm method.

| Category | Frequency | Percentage (%) | Acceptable rate |
|---|---|---|---|
| Exact | 295 | 91.61 | |
| Adequate | 12 | 3.73 | 95.34% |
| Inadequate | 15 | 4.66 | |

Table 2: Comparison between the AEPm method and previous methods.

| References | Acceptable rate (%) |
|---|---|
| Kwok *et al.* [10] | 83.9 |
| Raba *et al.* [38] | 86.0 |
| Li *et al.* [21] | 90.1 |
| Alam *et al.* [39] | 90.3 |
| Yoon *et al.* [27] | 92.2 |
| Yuksel *et al.* [28] | 94.4 |
| AEPm | 95.3 |

The adequate cases occurred mainly when the fibrous tissue of the breast overlapped the pectoral muscle, and therefore, the method could not recognize it exactly, but managed to recognize it with sufficient precision to locate the pectoral margin, as is shown in Figure 10(a) and 10(b).

In some cases where the pectoral muscle had very heterogeneous gray level values as is shown in Figure 10(c), the AEPm method can also found inadequate edges. In other images, the intensity of gray levels is very high and the difference between the pectoral region and breast tissue is minimal, and our method can not find the muscle edge, as is shown in Figure 10(d).

The *acceptable rate* as a method of evaluation may lead to discrepancies between the different authors, what means that an *"acceptable"* result can differ significantly between specialists. This method of evaluation is based on visual subjective opinion with little quantitative endorsement. Therefore, we also evaluated the AEPm method with a second criterion.

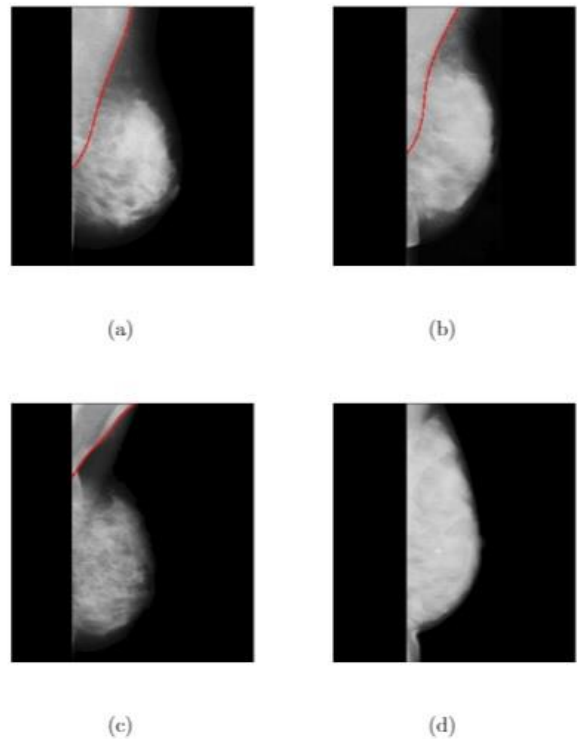

Figure 10: Examples of pectoral muscle edges adequately identified (a) mdb222 and (b) mdb242 in mini-MIAS. Examples of pectoral muscle edges inadequately identified or not found (c) mdb066 and (d) mdb318 in mini-MIAS.

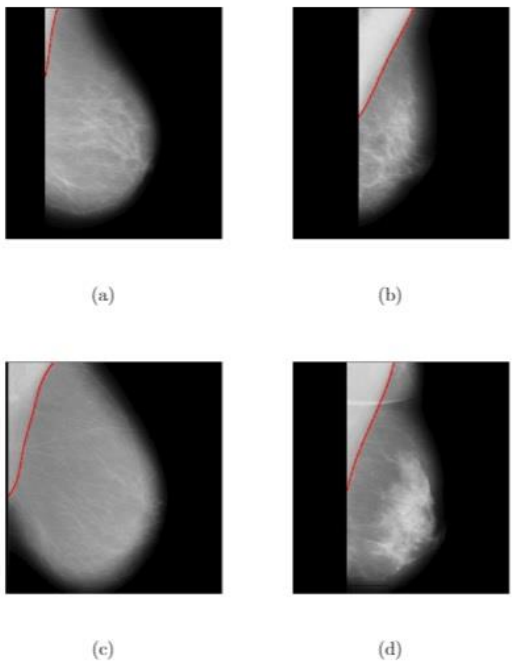

Figure 9: Examples of pectoral muscle edges exactly identified (a) mdb070, (b) mdb246, (c) mdb136 and (d) mdb161 in mini-MIAS.

## 3.1 Area normalized error

This method of evaluation is based on a particular set of data. The set contains 84 mammogram images from the mini-MIAS database and their coordinates of the pectoral muscle edge. These coordinates were marked by specialists and are taken as the real edges, (see [9]). The coordinates of the lines drawn by the radiologist in cited study were kindly provided by R. Ferrari.

The *area normalized error* is built with the percentages of false positive ($FP$) and false negative ($FN$) pixels. As is defined in Ferrari *et al.* [9], a $FP$ pixel is one included in the proposed region but not assigned in the reference region. Similarly, a $FN$ pixel is one assigned in the reference region but no included in the proposed region. The proportion of $FP$ pixels ($FP_I$) and proportion of $FN$ pixels ($FN_I$) for an image $I$ are computed by means of using the expression:

$$FP_I = \frac{1}{A(I)} \sum_{y=1}^{p} \max\{0, E_{pro}(y) - E_{ref}(y)\}$$

$$FN_I = \frac{1}{A(I)} \sum_{y=1}^{p} \max\{0, E_{ref}(y) - E_{pro}(y)\}, \quad (6)$$

where $A(I)$ is the area of the pectoral muscle in mammogram $I$ marked by the radiologists, $p$ is the number of rows in which the pectoral muscle appears in the mammogram, $E_{ref(y)}$ is the horizontal coordinate of the edge point in row $y$ drawn by the radiologists and $E_{pro(y)}$ is the coordinate of the edge point in row $y$ proposed by the method. The mean errors of $FP$ and $FN$ for the collection of mammograms $I$, $I = \{1, ..., N\}$ were calculated as:

$$FP_m = \frac{1}{N} \sum_{I=1}^{N} FP_I$$

$$FN_m = \frac{1}{N} \sum_{I=1}^{N} FN_I, \quad (7)$$

The performance analysis comparison using the *area normalized error* between the proposed method and other methods is given in Table 3.

In relation with the AEPm, the mean error of $FP$ was competitive with respect to the rest of the methods but the mean error of $FN$ was the best of the row. Moreover, in approximately 62% of the mammograms both errors were less than 0.05 and the fact of not having any mammogram in the higher error range shows a good performance for the proposed method.

Table 3: Performance analysis comparison using the area normalized error.

| | Hough [9] | Gabor [9] | A.P. [11] | Graph cut [15] | I.S. [21] | AEPm method |
|---|---|---|---|---|---|---|
| $FP_m$ | 0.0198 | 0.0058 | 0.0371 | 0.0064 | 0.0145 | 0.0196 |
| $FN_m$ | 0.02519 | 0.0577 | 0.0595 | 0.0558 | 0.0552 | 0.0486 |
| Number of image with: $(FP, FN) < 0.05$ | 10 | 45 | 50 | 43 | 48 | 52 |
| $\min(FP, FN) < 0.05, \max(FP, FN) < 0.10$ | 0 | 0 | 18 | 19 | 28 | 18 |
| $\min(FP, FN) < 0.05, \max(FP, FN) > 0.10$ | 0 | 0 | 11 | 22 | 7 | 14 |
| $0.05 < (FP, FN) < 0.10$ | 8 | 22 | 0 | 0 | 0 | 0 |
| $0.05 < \min(FP, FN) < 0.10, \max(FP, FN) > 0.10$ | 0 | 0 | 5 | 0 | 1 | 0 |
| $(FP, FN) > 0.10$ | 66 | 17 | 0 | 0 | 0 | 0 |

## 4. Conclusions

We proposed a new method (AEPm) for pectoral muscle elimination in MLO view mammograms. This method eliminates automatically the pectoral muscle region, and can be considered as a preprocessing step for future work focused on mammograms analysis. Our method takes advantage of the important anatomical features of the pectoral muscle. The use of Beta distribution provides a novel view for intensity level transformation.

The proposed algorithm was tested on 322 digitized mammograms from the Mini-MIAS, with an acceptable rate of 95.34%. Using a set of 84 mammograms provided by R. Ferrari we found that the mean error of false positive of the proposed method was comparable with other methods, the mean error of false negative obtained was the best and the distribution of the error terms was considerably good.

The proposed method is quite robust and solid given the heterogeneity (in all possible characteristics of the image) of the database with which it has been worked. This fact allows it to be quite adequate to apply to any other mammography database.

As part of the future research, we are developing a deep study of the beta transformation in order to improve its performance. It would also be interesting to extend the methodology that has been introduced in this paper to other types of medical images where it is necessary to pre-process the image based on a previous cleaning or segmentation. Based on the good results that have been obtained in this paper, our most imminent step is to try to automatically detect abnormalities in the breast and proceed to classify them, which would be of great support to radiologists.

**Acknowledgments**

The authors acknowledge financial support from *Ministerio de Ciencia e Innovación* grant ECO2015-66593-P and from the Basic Science Project *Modelos con estructuras de dependencias II*, (CONACYT CB-2015-01-252996). We would also like to thank R.J. Ferrari for providing the coordinates of the lines of pectoral muscle used in this work. We would also like to thank the reviewers for their contribution to improve the quality of this work.

**Jairo A. Ayala-Godoy.** Obtained his B.A. in Mathematics from Universidad Industrial de Santander (Colombia) in 2006; his M.Sc. in Statistics Mathematics from Universidad de Puerto Rico at Mayagüez (Puerto Rico) in 2012 and his Ph.D. in Probability and Statistics from Centro de Investigación en Matemáticas, A.C. (Mexico) in 2018. He is Full Professor of Department of Basic Sciences and Engineering in Universidad del Caribe (Cancun, Mexico), since August 2017. He has been advisor of 6 B.A. students and also he has collaborated in some research projects. His research interests include big data applied to medicine, functional data analysis, machine learning, data mining and geographic information systems.

**Rosa E. Lillo.** Obtained her B.A. with Honors and her Ph.D. in Mathematics from Universidad Complutense (Madrid) in 1992 and 1996, respectively. Since 2010, she is Full Professor of Statistics and Operations Research in Universidad Carlos III de Madrid. She has been Vice-Dean for the Degree in Statistics and Business between 2005 and 2015, Head of the Statistics Department (2015-september 2018) and Head of the UC3M-Santander Big Data Institute (From September 2018). She has published more than 60 research papers, and has been advisor of 12 Ph.D. students and also has collaborated in a wide variety of research projects. The intend of some of these projects is to propose solutions to problems faced by companies or public institutions. She obtained a Young Researcher Award for Research Excellence (UC3M-Banco Santander) in 2012, and in the past, she was visiting professor in The University of Arizona, Tucson. Her research interests include big data applied to medicine, multivariate risk measures, functional data analysis, BMAP processes and their applications in finance and queue networks, stochastic ordering and reliability and GLM models in high dimension.

**Juan Romo**. Obtained his B.Sc. in Mathematics with Honors from Universidad Complutense de Madrid and his Ph.D. in Mathematics from Texas A&M University. Postdoc in City University of New York. He has been Associate Professor, both in Universidad Complutense de Madrid and Universidad Carlos III de Madrid, where he is Professor of Statistics. He has been Vice-President for Graduate Studies and Vice-President for Faculty and Departments in Universidad Carlos III de Madrid and he is its President since March, 2015. President of the Young European Research Universities Network (YERUN) since September, 2015. Vice-President of Consejo de Universidades since September 2018. Permanent Comission of *"Conferencia de Rectores de las Universidades Españolas (CRUE)"* since July 2018. Coauthor of four books, he has published articles in research journals on complex, high-dimensional and functional data analysis, big data, resampling, time series, extremes and outliers, with applications in genetics (microarrays and SNPs), financial data, networks and image analysis.